\documentstyle[12pt,amstex]{article} 
  \newtheorem{df}{Definition}[section]
  \newtheorem{th}[df]{Theorem}
  \newtheorem{lem}[df]{Lemma}
  
  \newtheorem{prop}[df]{Proposition}
  \newtheorem{cor}[df]{Corollary}
  
  \makeatletter
   
   \@addtoreset{equation}{section}
  \makeatother

\begin{document}
 \title{ Nonlinear Grassmann Sigma Models in Any Dimension \\
         and \\
         An Infinite Number of Conserved Currents }
 \author{ 
  Kazuyuki FUJII,\thanks{Department of Mathematics, 
  Yokohama City University, 
  Yokohama 236, 
  Japan, \endgraf 
  {\it E-mail address}: fujii{\char'100}yokohama-cu.ac.jp} \ 
  Yasushi HOMMA\thanks{Department of Mathematics, 
  Waseda University, 
  Tokyo 169, 
  Japan, \endgraf 
  {\it E-mail address}: 696m5121{\char'100}mn.waseda.ac.jp} \ and 
  Tatsuo SUZUKI\thanks{Department of Mathematics, 
  Waseda University, 
  Tokyo 169, 
  Japan, \endgraf 
  {\it E-mail address}: 695m5050{\char'100}mn.waseda.ac.jp}}
 \date{}
 \maketitle
\begin{abstract}
We first consider nonlinear Grassmann sigma models in any dimension and next construct their submodels. For these models we construct an infinite number of nontrivial conserved currents.

Our result is independent of time-space dimensions and, therfore, is a full generalization of that of authors (Alvarez, Ferreira and Guillen).

Our result also suggests that our method may be applied to other nonlinear sigma models such as chiral models, $G/H$ sigma models in any dimension.
\end{abstract}
 \section{Introduction}
Nonlinear  (Grassmann) sigma models in two dimensions are very interesting objects to study in the not only classical but also quantum point of view and we have a great many papers on this topics. See, for example, Zakrzewski \cite{Zak}, Mickelsson \cite{Mic} and their references.

But in the dimensions greater than two, we have in general no outstanding results in spite of much efforts. See, for example, \cite{Fuj1},\cite{Fuj2},\cite{FR1},\cite{FR2}.

Recently Alvarez, Ferreira and Guillen in \cite{AFG} proposed a new approach to higher dimensional ``integrable" theories. Instead of higher dimensional nonlinear sigma models themselves (these ones are of course non integrable), they considered their submodels to construct ``integrable" theories.

In fact, as a simple example, they considered $CP^1$-model in $(1+2)$-dimensions
$$
(1+|u|^2)\partial^{\mu}\partial_{\mu}u
  -2\bar{u}\partial^{\mu}u\partial_{\mu}u =0
$$
$$
\mbox{for} \quad u:M^{1+2} \rightarrow \bf C 
$$
and constructed a submodel
$$
\partial^{\mu}\partial_{\mu}u=0 \quad \mbox{and} \quad 
\partial^{\mu}u\partial_{\mu}u =0
$$
and an infinite number of nontrivial conserved currents for this model.

Soon after their results were reinforced and generalized by Fujii and Suzuki \cite{FS1},\cite{FS2} and Gianzo, Madsen and Guillen \cite{GMG}.

But if we consider the submodel more deeply, we find that there is no reason to restrict the submodel to three dimensions. Namely, we may consider a model
$$
\partial^{\mu}\partial_{\mu}u=0 \quad \mbox{and} \quad 
\partial^{\mu}u\partial_{\mu}u =0
$$
$$
\mbox{for} \quad u:M^{1+m} \rightarrow \bf C 
$$
in any dimension ($m \in \bf N$). This means a kind of universality of the submodel.

After thoroughgoing analysis of the paper \cite{AFG}, we found that their method developed there was, more or less, irrelevant to construct submodels and conserved currents. We of course admit that \cite{AFG} is important, suggestive and instructive to nonexperts in this field.

In this letter, we define submodels of nonlinear Grassmann sigma models in any dimension and construct an infinite number of nontrivial conserved currents. 

Our results is a full generalization of \cite{FS1},\cite{FS2} and \cite{GMG}. Our method compared to that of \cite{AFG} is very simple and easy to understand.

%%%%%%%%%%%%%%%%%%%%%%%%%%%%%%%%%%%%%%%%%%%%%%%%%%%%%%%%%%%%%%%%%%%%%%%%%
 \section{Mathematical Preliminaries $\cdots$ Grassmann Manifolds}
Let $M(m,n;{\bf C})$ be the set of $m \times n$ matrices over $\bf C$ and we set $M(n;{\bf C}) \equiv M(n,n;{\bf C})$ for simplicity. For a pair $(n,N)$ with $1 \le n<N$, we set $I$, $O$ as a unit matrix, a zero matrix in $M(n;{\bf C})$ and  $I'$, $O'$ as ones in $M(N-n;{\bf C})$ respectively.

We define a Grassmann manifold for the pair $(n,N)$ above as
\begin{equation}
 G_{n,N}({\bf C}) \equiv \{ P \in M(N;{\bf C})| P^2=P, P^{\dag}=P, \mbox{tr}P=n \}.
\end{equation}
Then it is well-known that
\begin{eqnarray}
 G_{n,N}({\bf C}) 
 &=&
 \left\{ U
          \left(
           \begin{array}{cc}
               I &    \\
                 & O' \\
            \end{array}
           \right)
            U^{\dag}|U \in U(N) \right\}  \label{eqn:1-2}\\
 &\cong& \displaystyle{\frac{U(N)}{U(n) \times U(N-n)}}.\label{eqn:1-3}
\end{eqnarray}
It is easy to see $\mbox{dim}_{\bf C}G_{n,N}({\bf C})=n(N-n)$ from (\ref{eqn:1-3}). In the case $n=1$, we usually write $G_{1,N}({\bf C})={\bf C}P^{N-1}$ and call it the complex projective space. It is well-known 
\begin{equation}
 {\bf C}P^{N-1} \cong \displaystyle{\frac{U(N)}{U(1) \times U(N-1)}}
                \cong S_{\bf C}^{N-1}/S^1.
\end{equation}
Moreover, in the case $N=2$,
\begin{equation}
 G_{1,2}({\bf C})={\bf C}P^1 \cong S_{\bf C}^{2}/S^1 \cong S^2.
\end{equation}

Next, let us introduce a local chart for $G_{n,N}({\bf C})$. For $Z \in M(N-n;{\bf C})$ a neighborhood of 
         $\left(
           \begin{array}{cc}
               I &    \\
                 & O' \\
            \end{array}
           \right) $
in $G_{n,N}({\bf C})$ is expressed as 
\begin{equation}
 P_0(Z)=
          \left(
           \begin{array}{cc}
               I & -Z^{\dag}  \\
               Z & I' \\
            \end{array}
           \right)  
          \left(
           \begin{array}{cc}
               I &    \\
                 & O' \\
            \end{array}
           \right)  
          \left(
           \begin{array}{cc}
               I & -Z^{\dag} \\
               Z & I' \\
            \end{array}
           \right)^{-1} .
\end{equation}
This is also written as 
\begin{equation}
 P_0(Z)=
 \left(
   \begin{array}{cc}
     (I+Z^{\dag}Z)^{-1} & Z^{\dag}(I'+Z Z^{\dag})^{-1}  \\
     Z(I+Z^{\dag}Z)^{-1} & (I'+Z Z^{\dag})^{-1} \\
   \end{array}
 \right) .
\end{equation}
We note here relations
\begin{equation}
 Z(I+Z^{\dag}Z)^{-1}=(I'+Z Z^{\dag})^{-1}Z,
\end{equation}
\begin{equation}
 (I'+Z Z^{\dag})^{-1}=I'-Z(I+Z^{\dag}Z)^{-1}Z^{\dag}.
 \label{eqn:1-9}
\end{equation}
As for these tools, see, for example, \cite{FKaSa}. Since any $P$ in $G_{n,N}({\bf C})$ is written as 
$P=U
    \left(
       \begin{array}{cc}
           I &    \\
             & O' \\
       \end{array}
    \right) U^{\dag}$
for some $U \in U(N)$ by (\ref{eqn:1-2}), an element of a neghborhood of $P$ in $G_{n,N}({\bf C})$ is expressed as
\begin{equation}
 P(Z)=U P_0(Z) U^{\dag}.
 \label{eqn:1-10}
\end{equation}
Then,
\begin{lem}
 we have easily
\begin{equation}
(i) \quad dP_0=
         \left(
           \begin{array}{cc}
               I & Z^{\dag}  \\
              -Z & I' \\
            \end{array}
         \right)^{-1}  
         \left(
           \begin{array}{cc}
                 & dZ^{\dag}  \\
              dZ &    \\
            \end{array}
         \right) 
         \left(
           \begin{array}{cc}
               I & -Z^{\dag}  \\
               Z & I' \\
            \end{array}
         \right)^{-1},
 \label{eqn:1-11}
\end{equation}
\begin{equation}
(ii) \quad [P_0,dP_0]=
         \left(
           \begin{array}{cc}
               I & Z^{\dag}  \\
              -Z & I' \\
            \end{array}
         \right)^{-1}  
         \left(
           \begin{array}{cc}
                 & dZ^{\dag}  \\
             -dZ &    \\
            \end{array}
         \right) 
         \left(
           \begin{array}{cc}
               I & -Z^{\dag}  \\
               Z & I' \\
            \end{array}
         \right)^{-1}.
\end{equation}
\end{lem}

%%%%%%%%%%%%%%%%%%%%%%%%%%%%%%%%%%%%%%%%%%%%%%%%%%
 \section{Nonlinear Grassmann Sigma Models and Submodels}
Let $M^{1+m}$ be a $(1+m)$-dimensional Minkowski space $(m \in {\bf N})$. We consider a nonlinear Grassmann sigma model in any dimension. Let the pair $(n,N)$ be fixed. The action is 
\begin{equation}
 \mbox{$\cal{A}$}(P) \equiv \frac12 \int d^{1+m} x \ 
                             \mbox{tr}\partial_{\mu}P\partial^{\mu}P
 \label{eqn:2-1}
\end{equation}
where
$$
 P:M^{1+m} \longrightarrow G_{n,N}({\bf C}).
$$
Its equations of motion read 
\begin{equation}
 [P, \Box P] \equiv [P, \partial_{\mu}\partial^{\mu}P]=0.
 \label{eqn:2-2}
\end{equation}
From this
\begin{equation}
 0=[P, \partial_{\mu}\partial^{\mu}P]
  =\partial^{\mu}[P, \partial_{\mu}P],
\end{equation}
so $[P, \partial_{\mu}P]$ are conserved currents (Noether currents).

Next, we look for a local form of our action. By (\ref{eqn:2-1}) and (\ref{eqn:1-10}),(\ref{eqn:1-11}) we can put
\begin{equation}
 P(Z)=U
         \left(
           \begin{array}{cc}
               I & -Z^{\dag}  \\
               Z & I' \\
            \end{array}
         \right)  
         \left(
           \begin{array}{cc}
               I &     \\
                 &  O' \\
            \end{array}
         \right) 
         \left(
           \begin{array}{cc}
               I & -Z^{\dag}  \\
               Z & I' \\
            \end{array}
         \right)^{-1}  U^{\dag},
 \label{eqn:2-4}
\end{equation}
\begin{equation}
 \partial_{\mu}P(Z)=U
         \left(
           \begin{array}{cc}
               I & Z^{\dag}  \\
              -Z & I' \\
            \end{array}
         \right)^{-1}  
         \left(
           \begin{array}{cc}
                              & \partial_{\mu}Z^{\dag}  \\
              \partial_{\mu}Z &                         \\
            \end{array}
         \right) 
         \left(
           \begin{array}{cc}
               I & -Z^{\dag}  \\
               Z & I' \\
            \end{array}
         \right)^{-1} U^{\dag},
\end{equation}
where
$$
Z:M^{1+m} \longrightarrow M(N-n,n;{\bf C})
$$
and $U$ is a constant unitary matrix. Then,
\begin{lem}
we have \\
(i) action
\begin{equation}
 \mbox{$\cal{A}$}(Z)
  =\int d^{1+m} x \ 
    \mbox{tr}(I+Z^{\dag}Z)^{-1}\partial^{\mu}Z^{\dag}
             (I'+Z Z^{\dag})^{-1}\partial_{\mu}Z,
\end{equation}
(ii) the equations of motion
\begin{equation}
 \partial^{\mu}\partial_{\mu}Z
 -2\partial^{\mu}Z(I+Z^{\dag}Z)^{-1}Z^{\dag}\partial_{\mu}Z=0.
\end{equation}
\end{lem}
Let us consider the case $n=1$ (${\bf C}P^{N-1}$-model). If we set $Z=(u_1,\cdots ,u_{N-1})^t$ where $u_j:M^{1+m} \longrightarrow {\bf C}$ and remark that
$$
1+u^{\dag}u=1+\sum_{j=1}^{N-1}|u_j|^2,
$$
$$
(I'+u u^{\dag})^{-1}=I'-\frac{u u^{\dag}}{1+u^{\dag}u}
$$
from (\ref{eqn:1-9}),
\begin{cor}
we have \\
(i) action
\begin{equation}
 \mbox{$\cal{A}$}(u)
  =\int d^{1+m} x 
    \frac{(1+u^{\dag}u)\partial^{\mu}u^{\dag}\partial_{\mu}u-
          \partial^{\mu}u^{\dag}u u^{\dag}\partial_{\mu}u}{
           (1+u^{\dag}u)^2},
\end{equation}
(ii) the equations of motion
\begin{equation}
 (1+u^{\dag}u)\partial^{\mu}\partial_{\mu}u
 -2u^{\dag}\partial_{\mu}u\partial^{\mu}u=0.
\end{equation}
\end{cor}
These formulas are more familiar with us. Moreover in the case $N=2$ (${\bf C}P^1$-model),
\begin{cor}
we have \\
(i) action
\begin{equation}
 \mbox{$\cal{A}$}(u)
  =\int d^{1+m} x 
    \frac{\partial^{\mu}\bar{u}\partial_{\mu}u}{
           (1+|u|^2)^2},
 \label{eqn:2-10}
\end{equation}
(ii) the equations of motion
\begin{equation}
 (1+|u|^2)\partial^{\mu}\partial_{\mu}u
 -2\bar{u}\partial_{\mu}u\partial^{\mu}u=0.
 \label{eqn:2-11}
\end{equation}
\end{cor}
See \cite{AFG},\cite{FS1}.

Next, we define a submodel $\cdots$ a terminology of \cite{AFG} $\cdots$ of our model. Let us remind equations of motion of the model
$$
[P,\Box P]=0 .
$$
Since the tensor product $P \otimes P$ of $P$ is also projector, we assume the equations 
\begin{equation}
 [P \otimes P,\Box (P \otimes P)]=0. 
 \label{eqn:2-12}
\end{equation}
Transforming this, we have
\begin{equation}
 [P,\Box P] \otimes P + P \otimes [P,\Box P]
 +[P,\partial_{\mu}P] \otimes \partial^{\mu}P 
 +\partial^{\mu}P \otimes [P,\partial_{\mu}P]=0. 
\end{equation}
Now, let us define our submodel.
\begin{df}
We call a nonlinear Grassmann sigma model based on $P$ satisfying the simultaneous equations
\begin{equation}
 [P,\Box P]=0,
 \label{eqn:2-14}
\end{equation}
\begin{equation}
 [P,\partial_{\mu}P] \otimes \partial^{\mu}P 
 +\partial^{\mu}P \otimes [P,\partial_{\mu}P]=0
 \label{eqn:2-15}
\end{equation}
a submodel of this.
\end{df}
Note that (\ref{eqn:2-14}) and (\ref{eqn:2-15}) are equivalent to (\ref{eqn:2-2}) and (\ref{eqn:2-12}). 

Next, we want to express our submodel with $Z=(z_{ij})$ in (\ref{eqn:2-4}).
\begin{prop}
The equations above are equivalent to
\begin{equation}
 \partial^{\mu}\partial_{\mu}Z=0 \quad \mbox{and} \quad 
 \partial^{\mu}Z \otimes \partial_{\mu}Z=0
 \label{eqn:2-16}
\end{equation}
or in each component
\begin{equation}
 \partial^{\mu}\partial_{\mu}z_{ij}=0 \quad \mbox{and} \quad 
 \partial^{\mu}z_{ij}\partial_{\mu}z_{kl}=0
 \label{eqn:2-17}
\end{equation}
for any $1 \le i,k \le N-n, \ 1 \le j,l \le n$. 
\end{prop}
In this case $n=1, N=2$ (${\bf C}P^1$-model), we have
\begin{equation}
 \partial^{\mu}\partial_{\mu}u=0 \quad \mbox{and} \quad 
 \partial^{\mu}u\partial_{\mu}u=0
\end{equation}
with $u$ in (\ref{eqn:2-10}),(\ref{eqn:2-11}). This is a further generalization of that of \cite{AFG} because that is restricted to three dimensions. 

%%%%%%%%%%%%%%%%%%%%%%%%%%%%%%%%%%%%%%%%%%%%%%%%%%%%%%%%
 \section{An Infinite Number of Conserved Currents in Submodels}
It is usually not easy to construct conserved currents except for Noether ones in the nonlinear Grassmann sigma models in any dimension, but in our submodels we can easily construct an infinite number of conserved currents. This is a feature typical of our submodels. 

Our equations of submodel are
$$
 [P,\Box P]=0,
$$
$$
 [P,\partial_{\mu}P] \otimes \partial^{\mu}P 
 +\partial^{\mu}P \otimes [P,\partial_{\mu}P]=0
$$
in the global form (\ref{eqn:2-14}),(\ref{eqn:2-15}). Then, 
\begin{th}
for $j \geq 1$
\begin{equation}
 \tilde{B}_{\mu}^j \equiv 
   \sum_{k=0}^{j-1} 
    \underbrace{P \otimes \cdots \otimes P}_k
      \otimes [P,\partial_{\mu}P] \otimes 
    \underbrace{P \otimes \cdots \otimes P}_{j-1-k}
\end{equation}
are conserved currents:
\begin{equation}
 \partial^{\mu}\tilde{B}_{\mu}^j=0.
\end{equation}
\end{th}
For example, 
$$
\tilde{B}_{\mu}^3 = [P,\partial_{\mu}P] \otimes P \otimes P
                   +P \otimes [P,\partial_{\mu}P] \otimes P
                   +P \otimes P \otimes [P,\partial_{\mu}P],
$$
etc. Each component of $\tilde{B}_{\mu}^j$ is conserved currents, so we constructed an infinite number of conserved ones.

In particular, in the case $n=1, N=2$ (${\bf C}P^1$-model), let us write down the first column of $\tilde{B}_{\mu}^j$ which is the essential part. In this case
\begin{equation}
 P=\frac{1}{1+|u|^2}
   \left(
     \begin{array}{cc}
        1 & \bar{u} \\
        u & |u|^2 \\
     \end{array}
   \right),  
\end{equation}
\begin{equation}
 [P,\partial_{\mu}P]=\frac{1}{(1+|u|^2)^2}
   \left(
     \begin{array}{cc}
        \partial_{\mu}u\bar{u}-u\partial_{\mu}\bar{u}
             & \partial_{\mu}\bar{u}+\bar{u}^2 \partial_{\mu}u \\
        -(\partial_{\mu}u+u^2 \partial_{\mu}\bar{u})
             & -(\partial_{\mu}u\bar{u}-u\partial_{\mu}\bar{u}) \\
     \end{array}
   \right).
\end{equation}
\begin{cor}
For $j \geq 1$, 
\begin{equation}
 \tilde{B}_{\mu :k}^j=\frac{1}{(1+|u|^2)^{j+1}}
    \left\{
     j(\partial_{\mu}u\bar{u}-u\partial_{\mu}\bar{u})u^k
     -k(1+|u|^2)u^{k-1}\partial_{\mu}u
    \right\}
\end{equation}
and its complex conjugate are conserved currents, where $0 \le k \le j$ and we put $u^{-1}=0$ for $k=0$.
\end{cor}
This result recovers and, moreover, is simpler than that of \cite{FS1},\cite{FS2} and \cite{GMG}.

In the case $n=1$ (${\bf C}P^{N-1}$-model), the details of $\tilde{B}_{\mu}^j$ will be published in \cite{FHS}.

%%%%%%%%%%%%%%%%%%%%%%%%%%%%%%%%%%%%%%%%%%%%%%%%%%%%%%%%%%%%%%
 \section{A Class of Solutions in Submodels}
We cannot in general construct general solutions of nonlinear Grassmann sigma models in any dimension (see, \cite{FKoSa},\cite{FSa} and \cite{Zak} in two dimensions), but can construct a class of solutions in our submodels.

Our equations of submodels are
$$
 \partial^{\mu}\partial_{\mu}Z=0 \quad \mbox{and} \quad 
 \partial^{\mu}Z \otimes \partial_{\mu}Z=0
$$
or in each component
$$
 \partial^{\mu}\partial_{\mu}z_{ij}=0 \quad \mbox{and} \quad 
 \partial^{\mu}z_{ij}\partial_{\mu}z_{kl}=0
$$
in the local form (\ref{eqn:2-16}),(\ref{eqn:2-17}).
\begin{prop}
Let $f_{ij}$ ($1 \le i \le N-n, 1 \le j \le n$) be any function in $\mbox{C}\,^2$-class. Then
\begin{equation}
 z_{ij} \equiv f_{ij}(\alpha_0 t+\sum_{k=1}^m \alpha_k x_k)
\end{equation}
under
\begin{equation}
 \alpha^{\mu}\alpha_{\mu} \equiv \alpha_0^2-\sum_{k=1}^m \alpha_k^2=0
\end{equation}
is solutions of our submodels.
\end{prop}
Since $f_{ij}$ is any for $1 \le i \le N-n, 1 \le j \le n$, we constructed an infinite number of solutions of our models.

This situation is very similar to that of soliton theory.

%%%%%%%%%%%%%%%%%%%%%%%%%%%%%%%%%%%%%%%%%%%%%%%%%%%%%
 \section{Discussion}
We in this paper discussed the constructions of submodels of nonlinear Grassmann sigma models and of an infinite number of conserved currents.

Our result is a full generalization of that of \cite{AFG} and our method is much simpler than that.

Our discussion is restricted to Grassmann manifolds (these ones are easy to treat), but it can be generalized to other nonlinear sigma models whose target spaces are general symmetric spaces $G/H$ instead of Grassmann ones. This is now under study.

The detail and further developments of our result are published in \cite{FHS}.

 \section*{Acknowledgements}
Kazuyuki Fujii was partially supported by Grant-in-Aid for Scientific Research (C) No. 10640210. KF is very grateful to Prof. Akira Asada for useful suggestions. Yasushi Homma and Tatsuo Suzuki are very grateful to Tosiaki Kori for valuable discussions. 

%reference
  

\begin{thebibliography}{99}
  \bibitem{Zak}W. J. Zakrzewski:
  \newblock {\em Low Dimensional Sigma Models,}
  \newblock Adam Hilger, 1989.
%
  \bibitem{Mic}J. Mickelsson:
  \newblock {\em Current Algebras and Groups,}
  \newblock Plenum Press, 1989.
%
  \bibitem{Fuj1}K. Fujii:
  \newblock {\em Ferretti-Rajeev Term and Homotopy Theory,}
  \newblock Commun. Math. Phys., 162(1994), 273-287.
%
  \bibitem{Fuj2}K. Fujii:
  \newblock {\em Generalizations of the Wess-Zumino-Witten model and Ferretti-Rajeev model to any dimension: CP-invariance and geometry,}
  \newblock Jour. Math. Phys., 36(1995), 97-114.
%
  \bibitem{FR1}G. Ferretti and S. G. Rajeev:
  \newblock {\em Current Algebra in Three Dimensions,}
  \newblock Phys. Rev. Lett., 69(1992), 2033-2036.
%
  \bibitem{FR2}G. Ferretti and S. G. Rajeev:
  \newblock {\em ${\bf C}P^{N-1}$ Model with a Chern-Simon Term,}
  \newblock Mod. Phys. Lett., A7(1992), 2087-2094.
%
  \bibitem{AFG}O. Alvarez, L. A. Ferreira and J. S. Guillen:
  \newblock {\em A New Approach to Integrable Theories in Any Dimension,}
  \newblock hep-th/9710147.
%
  \bibitem{FS1}K. Fujii and T. Suzuki:
  \newblock {\em Nonlinear Sigma Models in $(1+2)$-Dimensions and An Infinite Number of Conserved Currents,}
  \newblock hep-th/9802105.
%
  \bibitem{FS2}K. Fujii and T. Suzuki:
  \newblock {\em Some Useful Formulas in Nonlinear Sigma Models in $(1+2)$-Dimensions,}
  \newblock hep-th/9804004.
%
  \bibitem{GMG}D. Gianzo, J. O. Madsen and J. S. Guillen:
  \newblock {\em Integrable Chiral Theories in 2+1 Dimensions,}
  \newblock hep-th/9805094.
%
  \bibitem{FHS}K. Fujii, Y. Homma and T. Suzuki:
  \newblock in preparation.
%
  \bibitem{FKaSa}K. Fujii, T. Kashiwa and S. Sakoda:
  \newblock {\em Coherent States over Grassmann Manifolds and the WKB Exactness in Path Integral,}
  \newblock Jour. Math. Phys., 37(1996), 567-602.
%
  \bibitem{FKoSa}K. Fujii, T. Koikawa and R. Sasaki:
  \newblock {\em Classical Solutions for the Supersymmetric Grassmannian Sigma Models in Two Dimensions I,}
  \newblock Prog. Theor. Phys., 71(1984), 388-394.
%
  \bibitem{FSa}K. Fujii and R. Sasaki:
  \newblock {\em Classical Solutions for the Supersymmetric Grassmannian Sigma Models in Two Dimensions II,}
  \newblock Prog. Theor. Phys., 71(1984), 831-839.
%
 \end{thebibliography}
\end{document}